\documentclass[12pt,preprint]{aastex}
\usepackage{graphicx}

\newcommand {\ea} {{\it et~al.}}
\newcommand {\be} {\begin{equation}}
\newcommand {\ee} {\end{equation}}

\shorttitle{Soft X-ray precursors}
\shortauthors{Moderski \ea}

\begin{document}

\title{Soft X-ray precursors of the non-thermal flares in blazars - 
theoretical predictions}

\author{Rafa{\l}~Moderski\altaffilmark{1}, 
Marek~Sikora\altaffilmark{1,2}, 
Greg~M. Madejski\altaffilmark{2}, 
and Tuneyoshi~Kamae\altaffilmark{2}}

\altaffiltext{1}{Nicolaus Copernicus Astronomical Center, Bartycka 18,
00-716 Warsaw, Poland; \\ \tt{moderski@camk.edu.pl}}
\altaffiltext{2}{Stanford Linear Accelerator Center, 2575 Sand Hill
Road, Menlo Park, CA 94025, USA}

\begin{abstract}
  Popular internal shock models, developed to explain production of
high energy flares in blazar jets, involve collisions between local
overdensities of matter being ejected by a central engine and moving
along the jet with different velocities.  Prior to such collisions,
the matter is relatively cold and therefore does not produce intrinsic
non-thermal radiation.  However, due to Comptonization of external
radiation by cold electrons, the presence of such matter should be
apparent by prominent precursor soft X-ray flares, visible prior to
non-thermal $\gamma$-ray flares.  In this paper we discuss the
predicted properties of such precursors and study the dependence of
their properties (luminosities and light curves) on kinematic
parameters of relativistic ejecta and on an angle of view.  We
demonstrate that the lack of evidence for luminous soft X-ray
precursors can be reconciled with our predictions for their properties
if acceleration and collimation of a jet takes about three distance
decades.  We briefly discuss the severe constraints on the internal
shock models that would be imposed by a non-detection of such
precursors.
\end{abstract}

\section{INTRODUCTION}
Strong jets in radio loud quasars must be launched very deeply in the
gravitational potential well, and most likely are powered by rotating
black hole and/or innermost parts of an accretion disk.  Their
formation is very likely mediated by magnetic fields, but the detailed
model is not yet established \citep[compare, e.g., scenarios suggested
by][]{con95,lnr97,vk03}.  In magnetic models, acceleration and
collimation processes are quite smooth and can proceed over several
distance decades.  It is still not known whether this is accomplished
on sub-parsec, parsec, or even larger scales.  \citet{vk03} claim that
some VLBI data on the variability of the apparent structure of parsec
scale jets are consistent with the acceleration process in the
Poynting flux dominated flow.  Those authors consider also circular
polarization \citep{haw01,ens03} and rotation-measure gradients
\citep{amy02,gm03} to be properties that strongly support dynamical
dominance of magnetic fields in parsec scale jets. However, all these
features don't have a unique interpretation: accelerating radio
features, which are actually observed very rarely, can represent
shocks formed on the interface between a jet and a clump of matter
entering the jet from outside and being accelerated by the
relativistic flow.  Likewise, the circular polarization does not
necessary require dynamically dominated helical magnetic fields
\citep{rb02}.  Finally, the gradients in the rotation measure can
result from a non-uniform distribution in the external Faraday screen
\citep{homan04}.

Independent means to verify whether quasar jets are Poynting flux or
matter dominated is by studying radiative properties of quasars with
jets oriented close to the line of sight. These quasars together with
BL Lac objects form the sub-class of active galactic nuclei (AGNs)
called {\it blazars}. In such objects, any radiation produced within
the jet is Doppler boosted into the direction of the observer and
dominates over other radiative components of the AGN.  As it was
pointed out by \citet{bs87} and explored by \citet{sm00}, propagation
of relativistic jets through very dense radiation fields present in
the quasar cores should result in a production of prominent spectral
``bumps'' in the soft X-ray band due to Comptonization of external UV
radiation by cold electrons in a jet.  The lack of such features in
the observed blazar spectra indicates that the acceleration and
collimation of a jet takes place at a distance $\sim 10^3$
gravitational radii from the central supermassive black hole.  This
strongly supports the idea that jets in quasars are, at least
initially, magnetically dominated.

The question whether the jets are still magnetically dominated at
distances $\sim 0.1-3$ pc, where most of blazar radiation is produced,
remains open.  The most popular blazar models invoke internal shocks,
assumed to be produced by inhomogeneities moving down the jet with
different velocities \citep{sbr94,sglc01}.  Since efficient, strong
shocks cannot be formed in magnetically dominated jets, in these
models the energy flux of relativistic flow is dominated by matter,
specifically by protons. Such jets are likely to contain also
e$^+$e$^-$ pairs, but with the number $n_e/n_p < $tens (as can be
inferred from the shock energetics; see Appendix~\ref{appA1} and
\ref{appA4}), they contribute to the inertia of a jet only negligibly.
Prior to their collision, the inhomogeneities are cold, and, as for
the case of the cold stationary jet, they upscatter external UV
photons up to soft X-ray energies, forming X-ray ``precursors'' which
should precede the actual non-thermal flares.  Hence, a comparison of
the observational data against detailed predictions about the light
curves of the soft X-ray precursors and the non-thermal flares can
provide an efficient tool to search for an evidence for
inhomogeneities, the basic ingredient of the internal shock models
\citep{msk03,sm02}.  If such precursors are not detected, it may
indicate that non-thermal flares are produced by reconnection events
rather than by shocks and, therefore, that jets at parsec-scale
distances are still dominated by magnetic fields.  At the very least,
this can provide the lower limit to the distance where the jets
undergo a conversion from the magnetic to matter dominated flows and
where shocks could be products of the reconnection events.

In this paper, we develop theoretical framework describing the
properties of soft X-ray precursors.  We show the dependence of their
luminosities on kinematical parameters of ejecta and on the viewing
angle, and illustrate how their light curves should compare with the
time profiles (including time leads/lags) of the non-thermal flares.

\section{ASSUMPTIONS AND EQUATIONS}
In our studies of X-ray precursors, we approximate the inhomogeneities
by shells which propagate ballistically, within boundaries of a cone.
In order to illustrate in a simple analytical manner the basic
relations between the properties of precursors and non-thermal flares,
we assume that the shells have equal masses and proper widths, denoted
as $\lambda_0$.  We also assume that the conical angle, $\theta_j$, is
sufficiently small that the kinematics and dynamics of ejecta can be
described using a one-dimensional model. With the former assumption,
the efficiency of the energy dissipation in the colliding shells is
maximized, and, therefore, the magnitude of the bulk-Compton effects
will be somewhat underestimated. Regarding the second assumption: the
one-dimensional approach can directly be applied for models with very
narrow jets and, in numerical models, it is generalized for a wider
jets by summing radiative contributions from the $\delta \theta \ll
\theta_j$ shell segments.

\subsection{Bulk-Compton luminosity}
Cold electrons, carried by two neighboring relativistic shells through
the external diffuse radiation field with their respective bulk
Lorentz factors $\Gamma_2 > \Gamma_1 \gg 1$, upscatter the ambient
photons with energy $h\nu_{diff}$ up to energies
\be
h\nu_{BC} \simeq \Gamma_i {\cal D}_i h\nu_{diff} \, ~~~ i=1,2 \, ,
\ee
where
\be
{\cal D}_i \equiv {1 \over \Gamma_i (1 - \beta_i \cos \theta_{obs})} \, ,
\ee
and $\theta_{obs}$ is the angle between the line of sight and the jet
axis.  The scattered photons form the so-called {\it bulk Compton
radiation}.  That radiation is beamed into the jet direction giving the
isotropic luminosity 
\be
L_{BC,i} \equiv 4 \pi {\partial L_{BC,i} \over \partial
\Omega_{\vec n_{obs}}} \simeq \int_{r_0}^{r_{sh}} 
\left(\partial N_{e,i} \over \partial r \right)_{obs}
{dE_{sc,i}' \over dt'} {\cal D}_i^4  \, dr   
\,
\label{LBC}
\ee
where $r_0$ is the distance at which the jet is fully formed and accelerated;
$r_{sh}$ is the distance of the discontinuity surface
which separates fluids of the two colliding shells;
\be 
{dE_{sc,i}' \over dt'} = {dE_{sc,i} \over dt} = 
{4 \over 3} \Gamma_i^2 c \sigma_T u_{diff} \, 
\label{DOTE}
\ee
is the amount of the scattered radiation energy per unit time, where
$u_{diff}$ is the energy density of the external radiation field; and
$(\partial N_{e,i}/\partial r)_{obs}dr$ is the number of electrons
contributing to the observed radiation at a given instant from a
distance range $dr$.  For a uniform radial distribution of electrons
within the range $r_{rear}^{(obs)} < r < r_{front}^{(obs)}$
\be
\left(\partial N_{e,i} \over \partial r \right)_{obs} = 
{N_{e}  \over \lambda_0 {\cal D}_i} \, ,
\label{dNdr}
\ee
and otherwise is zero.  Note that the above is due to the fact that
the source is observed as being stretched by a Doppler factor,
i.e. that $r_{front}^{(obs)}-r_{rear}^{(obs)}=\lambda_0 {\cal D}_i$.

Using Eqs. (\ref{LBC})-(\ref{dNdr}), one can find that for
$u_{diff}={\rm const}$ up to a given distance $r_m$ and then falling
very rapidly beyond $r_m$, the maximum bulk Compton luminosity,
$L_{BC}$, is
\be
L_{BC,i} \simeq 
{4 \over 3} c \sigma_T u_{diff} \Gamma_i^2 {\cal D}_i^4 N_e \epsilon_i \, ,
\label{Lmax}
\ee  
where 
\be
\epsilon_i = {\rm Min}[1;{r_m \over \lambda_0 {\cal D}_i}] \, .
\label{epsilon}
\ee

\subsection{Number of electrons}
\label{numele}
The number of cold electrons carried by a pair of shells is equal to
the number of relativistic electrons involved in the production of
non-thermal flares.  Assuming that relativistic electrons are injected
with a single power law distribution, $Q = K \gamma^{-p}$, the number
of relativistic electrons injected during the collision time scale,
$t_{coll}'= {\cal D}t_{fl}$, is
\be 
N_{e,inj} = t_{fl} {\cal D} \int_{\gamma_{min}}^{\gamma_{max}} Q \, d\gamma
= { t_{fl} {\cal D} K\over (p-1) \gamma_{min}^{p-1}}= 2 N_e \,
\label{Ne}
\ee
where $t_{fl}$ is the time scale of the non-thermal flare and $p>1$.

The normalization factor $K$ of the injection function $Q$ can be
found from the formula for emissivity which in the $\gamma$-ray band
is very likely dominated by Comptonization of external diffuse
radiation fields by ultra-relativistic electrons \citep[see, e.g.,
review by][]{sm01}.  At a given frequency we have
\be
\nu L_{\nu} \simeq {3 \over 8} {{\cal D}^6 \over \Gamma^2}
[N_{\gamma} \gamma] \, |\dot \gamma|_{EC} m_e c^2 \,
\label{ELE} ,
\ee
where $N_{\gamma} \equiv \partial N_e/\partial \gamma$ is the
differential energy distribution of electrons and $|\dot\gamma|_{EC}$
is the rate of the radiative losses of electrons due to the external
Compton process.

Radiation observed in the MeV-GeV band is produced by electrons which
cool on time scales much shorter than the dynamical time scale
(i.e. in the so-called fast cooling regime). For them, $N_{\gamma}
\simeq \int_{\gamma} Q d\gamma / |\dot \gamma|_{EC}$, and inserting
this into the Eq.~(\ref{ELE}) gives
\be
K= {8(p-1) \over 3 m_e c^2 \gamma_E^{-p+2}} (\nu_E L_{\nu_E})
{\Gamma^2 \over {\cal D}^6} \, ,
\label{K}
\ee
where $\gamma_E \simeq \sqrt{\nu_E/\nu_{diff}} / {\cal D} $ and
$\nu_E$ is the frequency at which the $\gamma$-ray luminosity is
measured.

For $p \simeq 2$, which corresponds with the typical value of the
energy flux spectral index in the EGRET band $\alpha \sim 1$
\citep{phjs97},
\be
K \simeq {8(\nu_E L_{\nu_E}) \over 3 m_e c^2} {\Gamma^2 \over {\cal
D}^6} \, ,
\label{K1}
\ee
and using Eq. (\ref{Ne}) we obtain 
\be
N_e \simeq {{\cal D} t_{fl} K \over 2 \gamma_{min}} \simeq 
{4 t_{fl}(\nu_E L_{\nu_E}) \over 3 m_e c^2} 
{\Gamma^2 \over {\cal D}^5}{1 \over \gamma_{min}} \, , 
\label{Ne1}
\ee
where $\gamma_{min} \le $ a few, as indicated by power-law
X-ray spectra extending down to energies $\le 1$keV \citep{tmg00} and
by circular polarization measurements \citep{whor98}.

It is worthwhile to note that the number of electrons in cold shells
could, in general, be much lower than that implied by the non-thermal
flares, since a pair production via absorption of highest energy
$\gamma$-rays in the $\gamma\gamma \to e^+e^-$ process is always
possible in principle.  However, the external radiation Compton models
predict the $\gamma$-ray spectra of quasars having the intrinsic
cutoff at energies not much larger than $10$ GeV \citep[see,
e.g.][]{sbr94}.  This together with characteristic $\gamma$-ray
spectral slopes $\alpha \sim 1$ indicates that the number of pairs
produced in the shock cannot significantly exceed the number of the
primary electrons and positrons in the ejecta. It also should be
mentioned here that in general population of relativistic particles
in the shocked plasma can be accompanied by population of (quasi-)thermal 
particles. In such a case, the total number of electrons/positrons would 
be larger and the precursors would be more luminous than presented 
in \S3, leading to stronger constraints on the internal shock models discussed 
in \S4.  

\subsection{The shell width}
In the internal shock model, the time scale of the non-thermal flares
is related to the time scale of the shock operation. The latter is
equal to the time scale of collision which depends on the velocities
of the shells and on their widths. Since internal shocks are only
marginally relativistic, the shell width can be derived by using the
formula which ignores the width of the shocked plasma layer.
In the shell rest frame this width is
\be
\lambda_0 = \lambda_0' \Gamma_{1,2}'= c t_{fl} {\cal D}
|\beta_{1,2}'| \Gamma_{1,2}' \simeq c t_{fl} {\cal D} {\alpha_{\Gamma}-1
\over 2 \sqrt{\alpha_{\Gamma}}} \, ,
\label{lambda}
\ee
where $|\beta_{1,2}'| = \beta_2' = -\beta_1' = (\alpha_{\Gamma}
-1)/(\alpha_{\Gamma} +1)$, $\Gamma_{1,2}' = \Gamma_2' = \Gamma_1' =
(\alpha_{\Gamma} +1)/(2\sqrt{\alpha_{\Gamma}})$, and $\alpha_{\Gamma}
\equiv \Gamma_2/\Gamma_1$ (see Appendix~\ref{appA2}).  Comparison of
the approximate value of $\lambda_0$ given by Eq.~(\ref{lambda})
with its exact value is provided in Appendix~\ref{appA3}. It can be
verified using formula (\ref{ll}) that our approximation leads to an
overestimate of $\lambda_0$ by no more than a factor 1.4. We use the
approximate formula in order to calculate self-consistently the light
curves, which, in addition to the precursors, include the non-thermal
flares.  The latter are computed by the BLAZAR code \citep{msb03}
which uses thin shell approximation.

\section{RESULTS}
The EGRET instrument detected many prominent flares in blazars
\citep{mbc95}.  In the 30 MeV - 10 GeV band, their apparent
luminosities reach values $\nu L_{\nu} \sim 10^{48}$ erg s$^{-1}$, and
their time scales, $t_{fl}$, are on the order of a few days. Relating
$t_{fl}$ to the shock life-time, one can find that the distance range
of the shock operation is
\be
\Delta r_{sh} = {\beta c t_{fl} \over 1 -\beta \cos{\theta_{obs}}}
\equiv \beta c t_{fl} {\cal D} \Gamma \simeq 7.8 \times 10^{17} \left(
{t \over 3~ {\rm day}} \right) \left( {\Gamma \over 10} \right)^2 {\rm
cm} \, .
\label{drfl}
\ee
Roughly symmetrical profiles of the observed non-thermal flares 
suggest that they occur at distances $r_{sh} \sim \Delta r_{sh}$ 
\citep[see][]{sbbm01}. At such distances, the energy density of the 
external diffuse radiation field is dominated by broad emission lines (BEL).
From reverberation mapping of the BEL regions in AGNs and quasars we know
that these lines are produced at 
\be
r_{BEL} \sim 10^{18} \left( {L_{UV} \over 10^{46}~ {\rm erg \,
s}^{-1}} \right)^{0.7} {\rm cm} \, ,
\label{rbel}
\ee
\citep{kaspi00}.  

\begin{figure}
\centering
\includegraphics[width=5in,angle=0]{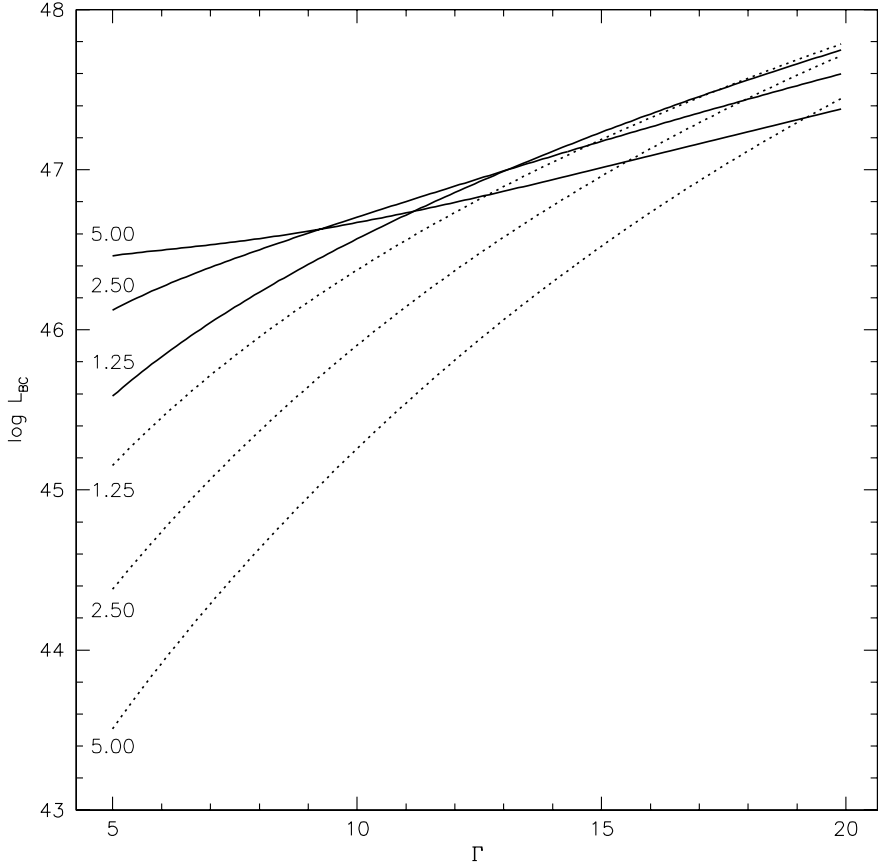}
\caption{ Precursor luminosity as a function of the bulk Lorentz
  factor $\Gamma$. The angle of view is $\theta_{obs} = 1/10$, and the
  three pairs of curves are calculated for ratio of the Lorentz
  factors of the two shells $\alpha_{\Gamma} = 1.25$; $2.5$; and $5$
  (marked beside the curves on the left side of the plot). The solid
  lines are for faster of the two shells, the dotted lines are for the
  slower of the two shells. The models are calculated for: $u_{diff}=
  {\rm const}=2.7 \times 10^{-3}$ erg cm$^{-3}$ for $r \le r_{BEL}$
  and $u_{diff}=0$ for $r > r_{BEL}$; $\nu_E L_{\nu_E} = 10^{48}$ erg 
  \, s$^{-1}$; $t_{fl} = 3$ d; and $\gamma_{min} =1$. Jet is assumed to be 
  very narrow ($\theta_j \ll 1/\Gamma_2$). }
\label{figa}
\end{figure}

\begin{figure}
\centering
\includegraphics[width=5in,angle=0]{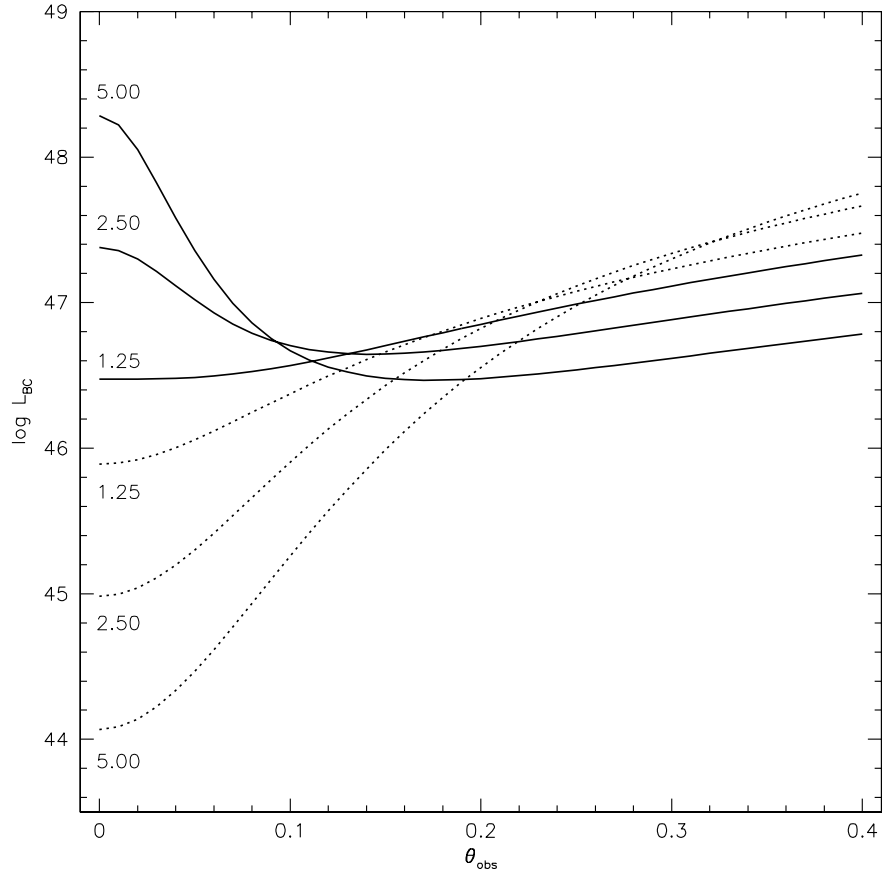}
\caption{
  Precursor luminosity as a function of the angle of view
  $\theta_{obs}$, for $\Gamma = 10$. Other parameters are the same as
  in Figure~\ref{figa}.}
\label{figb}
\end{figure}

Energy density of BEL radiation at $r<r_{BEL}$ is
\be u_{\rm BEL} \simeq {L_{\rm BEL} \over 4 \pi c r_{\rm BEL}^2}
\simeq 2.7 \times 10^{-3} \left({\xi_{\rm BEL} \over 0.1} \right)
\left( {L_{UV} \over 10^{46}~ {\rm erg \, s}^{-1}} \right)^{-0.4} {\rm
erg \, cm}^{-3} \, \ee
where $\xi_{\rm BEL} = L_{\rm BEL}/L_{\rm UV}$. At $r>r_{\rm BEL}$ the
energy density of BEL radiation drops faster than $r^{-2}$ and, therefore, in
the first approximation, the production of the bulk Compton radiation
at such distances can be ignored. Another effect which in the first
approximation can be neglected arises due to the decreasing number of
cold electrons due to the collision and shock formation.  With the
above approximations, one can estimate the maximum bulk Compton
luminosities using Eq. (\ref{Lmax}) for $r_m = r_{BEL}$.  The results
are presented in Fig.~\ref{figa}, which shows the dependence of the
precursor luminosities on the bulk Lorentz factor of the shocked
plasma, and in Fig.~\ref{figb}, which illustrates the dependence of
the luminosities on the angle of the jet with respect to our line of
sight.  In both Figures, we plot separately the value of $L_{BC}$ for
each of the two shells for various values of $\alpha_{\Gamma}$ (the
ratio of Lorentz factors of the two shells); both shells will
subsequently undergo a collision, which will result in the non-thermal
flare.  The values of $L_{BC}$ plotted there are calculated by
combining Eqs. (\ref{Lmax}), (\ref{epsilon}), (\ref{Ne1}), and
(\ref{lambda}), which give $L_{BC,i} \propto \Gamma_i^2 \Gamma^2 {\cal
D}_i^4/{\cal D}^5$ for $\epsilon_i=1$ (entire source is contributing
to the observed radiation) and $L_{BC,i} \propto \Gamma_i^2 \Gamma^2
{\cal D}_i^3/{\cal D}^6$ for $\epsilon_i<1$.

It is apparent from Fig.~\ref{figa} that the luminosity of the
precursors produced by the faster of the two shells becomes much
larger than that due to the slower shell when the bulk Lorentz factor
decreases.  The large difference at small $\Gamma$ factors reflects
the stronger Doppler boosting of radiation from the faster moving
sources than from the slower ones. The decrease of this difference
with increasing $\Gamma$ is caused mainly by a large reduction of
the beaming factor of the faster shell for the observer located at
fixed angle outside the Doppler cone.  

Fig.~\ref{figb} shows that bulk luminosities from faster moving shells
have lowest values at $\theta_{obs} \sim 1/\Gamma$.  For smaller
$\theta_{obs}$, the shape of $L_{BC,2}(\theta_{obs})$ scales roughly
with ${\cal D}_2^3$; for larger $\theta_{obs}$, the increase of bulk
Compton luminosity is mainly due to increase of a number of electrons
required to produced a given luminosity of non-thermal flares.  Since
for $\theta_{obs} < 1/\Gamma$ the dependence of ${\cal D}_1$ on
$\theta_{obs}$ is very flat, the shape of $L_{BC,1}(\theta_{obs})$ is
determined by a factor ${\cal D}^{-5}$ and, therefore, $L_{BC,1}$
increases with the viewing angle.  This increase continues at larger
angles, but is diminished due to the term ${\cal D}_1^4$, which starts
to contribute more strongly to the dependence of the $L_{BC,1}$ on the
angle of view for $\theta_{obs} > 1/\Gamma_1$.  Since the bulk Compton
luminosities presented on Figs.~\ref{figa} and \ref{figb} are
calculated for $\theta_j=0$, while real sub-parsec jets are expected
to have $\theta_j \sim 1/\Gamma$, the very high values of $L_{BC,2}$
at $\theta_{obs}<1/\Gamma$ are probably rather unrealistic.

\begin{figure}
\centering
\includegraphics[width=5in,angle=0]{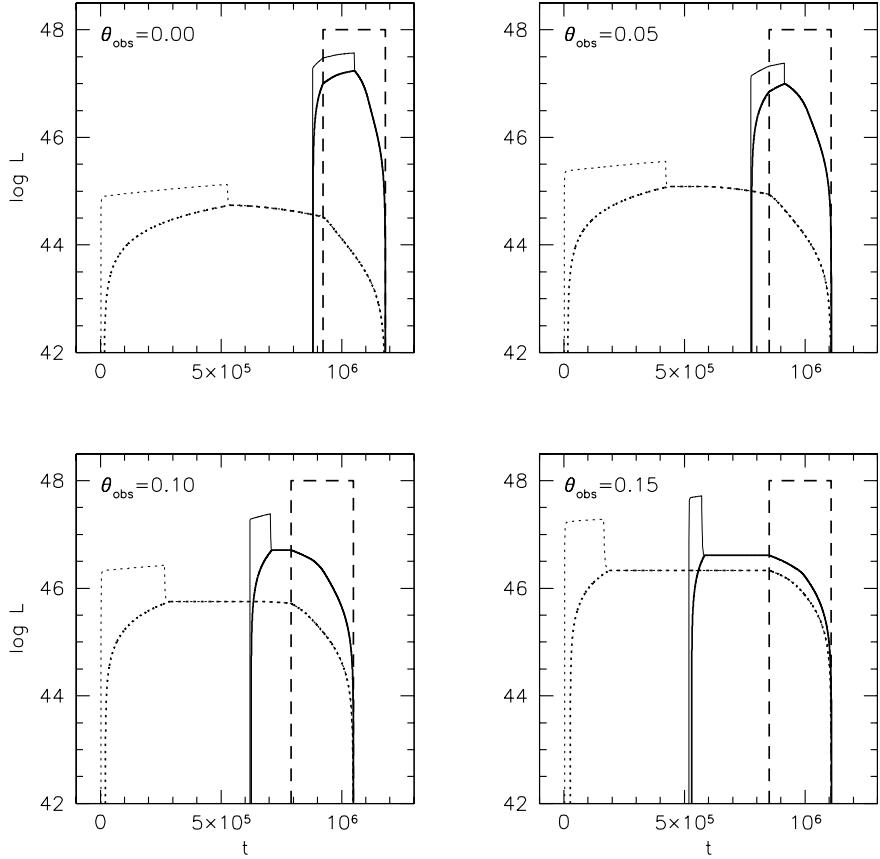}
\caption{ Light curves of precursors and flares 
  for four different values of $\theta_{obs}$.
  The solid lines are the precursors produced by the faster shells;
  the dotted lines are the precursors produced by the slower shells;
  and the dashed lines are the non-thermal $\gamma$-ray flares as
  produced deeply in the fast cooling regime.  The models are computed
  for $u_{diff}={\rm const}$ at $r_d < r < r_{BEL}$, and for the
  power-law distributions at $r<r_d$ and $r>r_{BEL}$ as described in
  the text.  Precursors starting at $r_0 = r_d = 3 \times
  10^{16}$ cm and $r_0 = 3 \times 10^{15}$ cm are shown by the thick and
  thin curves, respectively.
  Production of non-thermal flares by shocks starts at a distance
  $r_{0,sh}= 7.8 \times 10^{17}$ cm; the shocks terminate at a distance
  $r_{0,sh} + \Delta r_{sh}$ (see Eq. \ref{drfl}).  The kinematical
  parameters are $\Gamma=10$ and $\alpha_{\Gamma} = 3$.  Other
  parameters are the same as in Figures~\ref{figa} and \ref{figb}.}
\label{figc}
\end{figure}

\begin{figure}
\centering
\includegraphics[width=5in,angle=0]{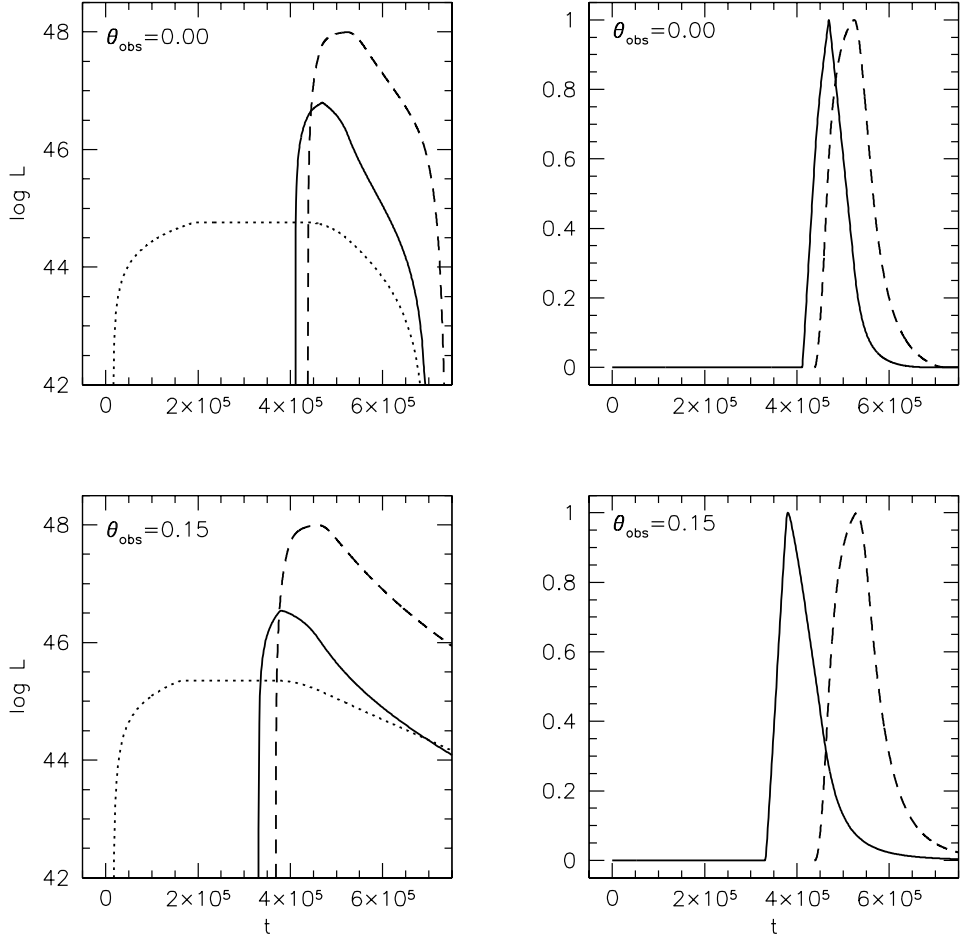}
\caption{
  Light-curves of the two types of precursors and the non-thermal
          flares for $\theta_{obs}=0$ (the upper left panel) and
          $\theta_{obs}=0.15$ (the lower left panel) for
          $\theta_j=1/\Gamma=0.1$.  Those due to the faster shells and
          the flares are re-drawn in linear scale on the right panels
          with their peaks normalized to one.  The models are computed
          for: $t_{fl}=1$d; $r_{0,sh}=\Delta r_{sh}$; and $r_0=r_d$.
          Other parameters are the same as in Figure~\ref{figc}.}
\label{figd}
\end{figure}

\begin{figure}
\centering
\includegraphics[width=5in,angle=0]{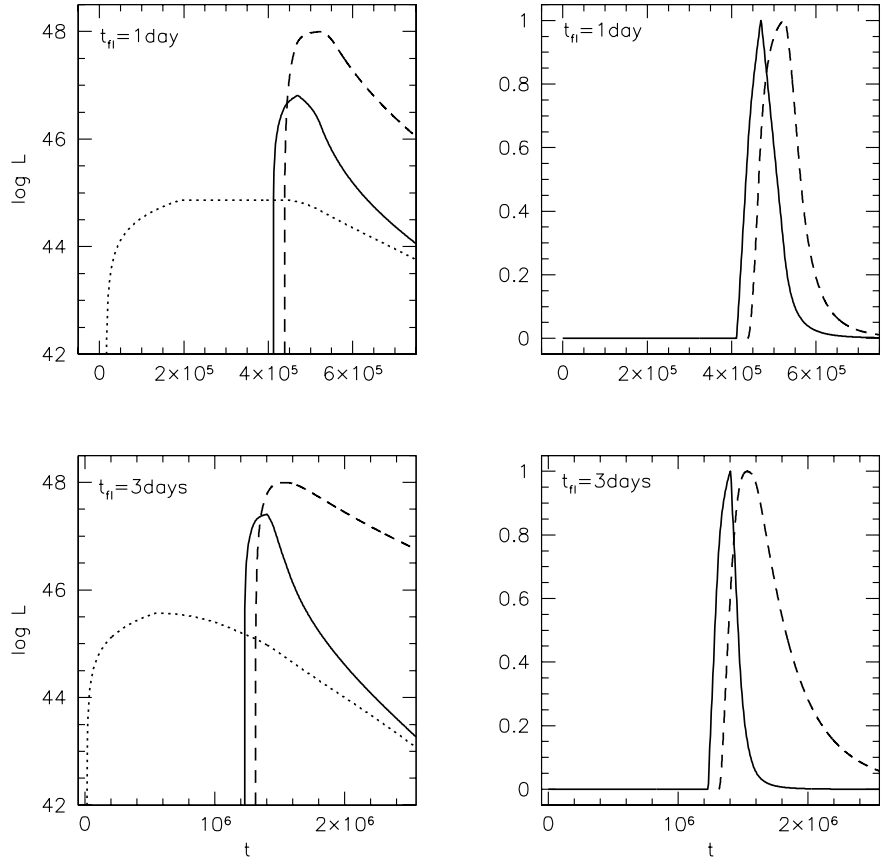}
\caption{
  Comparison of light-curves computed for $t_{fl}=1$ day and
  $3$ days.  Models are computed for $\theta_{obs}=1/\Gamma=0.1$.  All
  other parameters are the same as in Figure~\ref{figd}.}
\label{fige}
\end{figure}

The light-curves of soft X-ray precursors and of non-thermal flares
are presented in Fig.~\ref{figc} - for jets with $\theta_j \ll
\Gamma_2$, and in Fig.~\ref{figd} and \ref{fige} - for jets with
$\theta_j = \theta_{obs}= 1/\Gamma$.  These light curves have been
calculated for: $u_{diff} \propto r^{-3}$ for $r < r_d = 3 \times
10^{16}$ cm; $u_{diff} = u_{BEL}$ for $r_d < r < r_{BEL}$; and
$u_{diff} \sim r^{-2.5}$ for $r>r_{BEL}$.  The first branch mimics the
contribution of the seed photons for the EC process from the accretion
disk \citep{ds02}. This dominates radiative energy losses of
relativistic electrons at $r < r_d$.  The latter branch represents the
extension of BEL emissivities beyond the distance $r_{BEL}$.  The time
profiles of the non-thermal flares for $\theta_j = 1/\Gamma$ are
computed using the code BLAZAR described in \citet{msb03}.

It is apparent from Fig.~\ref{figc} and from the left panels of
Figs.~\ref{figd} and \ref{fige} that precursors produced by slower
shells are very extended in time and reach their maximum luminosities
significantly before the emergence of non-thermal flares. In contrast,
the brighter precursors, presumably produced by the faster shells,
overlap significantly with the non-thermal flares, and, for
$\theta_{obs} \le 1/\Gamma$, their peaks only marginally precede the
peaks of non-thermal flares.  (see right panels of Figs.~\ref{figd}
and \ref{fige}). More significant feature of the brighter precursors
is much faster decay of their light curves than of non-thermal flares.

In all Figures except for Fig.~\ref{figc}, the cold shells are assumed
to be formed at a distance $r_d$, above which energy density of
external radiation field as measured in a jet comoving frame is
dominated by the broad emission line radiation.  In Fig. \ref{figc},
the shells are assumed to be formed in the close proximity to the
central engine and the light-curves plotted in those Figures include
both Comptonization of broad emission lines and Comptonization of the
accretion disk radiation.  Comptonization of the disk radiation
produces additional ``bumps'' in the light curves of the precursors,
with a sharp drop at the time corresponding to $r_d$.  This sudden
drop is a result of the apparent lengths of the shells being stretched
by relativistic effects. For the parameters from Figure~\ref{figc} the
apparent length of the shell is much larger than the size of the
region in which radiation from the disk dominates, while the observed
travel time through the region is very short, since it is
relativistically reduced by a factor $(1 - \beta_i\cos\theta_{obs})$.
This results in a nearly rectangular shape of that part of the
precursor lightcurve.  Figure~\ref{figc} clearly shows that the
Comptonization of the disk radiation provides about three times larger
luminosity than Comptonization of broad emission lines, as indicated
by maximum amplitude of the precursors in both cases.

\section {DISCUSSION AND CONCLUSIONS}

If the $\gamma$-ray flares observed in many blazars are produced via
internal shocks, they should be accompanied by soft X-ray
precursors.  The precursors are predicted to result from Comptonization
of external UV continuum and broad emission line photons by cold
electrons carried by the relativistic ejecta.  The ejecta, moving with
different velocities, collide with each other and form shocks where
relativistic electrons are accelerated.  The dynamics of such events
is very complex, and depends on the unknown structure and kinematics
of the ejecta.  However, some basic properties, such as minimal
luminosities and profiles of the light curves of precursors vs. those
of non-thermal flares should not depend strongly on the details of the
model. We study them by approximating the ejecta by discrete,
intrinsically identical shells.  We analyze the dependence of precursors
on various parameters such as the bulk Lorentz factor of the shocked
plasma, $\Gamma$, the ratio of the bulk Lorentz factors of shells,
$\alpha_{\Gamma} = \Gamma_2/\Gamma_1$, and the viewing angle,
$\theta_{obs}$.

Our results show that precursors produced by faster shells are
typically 10 - 30 times less luminous than $\gamma$-ray flares if the
shells are formed at $r_0=r_d$, and about $10$ times less luminous if
the shells are formed much closer to the central engine and
contribution from the Comptonization of the disk radiation is
included.  We note here that our model predicts that the low energy
tail of the EC component of the non-thermal flare contributes in the
soft X-ray band about 30 - 100 less than at the peak \citep[see,
e.g.,][]{msb03}.  The Synchrotron Self-Compton (SSC) contribution is 
also predicted to be of the same order as EC component.  
With this, the precursors from the
faster shells are predicted to dominate in the soft X-ray band over
the non-thermal flares, even if the contribution from the accretion
disk is not included.

As shown in Figures~\ref{figd} and \ref{fige}, the brighter precursors
overlap significantly in time with non-thermal flares, but because
their flux should decay much faster than for the non-thermal flares,
they should be easily distinguished from the synchrotron and SSC
contributions, provided they are monitored with the intra-day time
resolution. Furthermore, presence of precursors can be verified by
future polarimetric measurements. As predicted by \citet{bs87}, the
bulk Compton radiation should be highly polarized, with the electric
vector perpendicular to the jet axis. The polarized soft X-ray
radiation is expected to be produced also by synchrotron and SSC
process.  However, for magnetic fields dominated by the transverse
component (due to compression of chaotic magnetic fields in the
transverse shocks), that radiation should be polarized with the
electric vector parallel to the jet axis. Hence, the polarization
position angle measurements and the variability analysis can provide
two independent diagnosis of the bulk-Compton radiation.

Clearly, search for the soft X-ray precursors is a viable avenue to
study the nature of blazar jets, and in particular, to determine if
their kinetic luminosities are dominated by energy flux of particles
or by Poynting flux. 
The lack of observational evidence for precursors by future observations
will indicate that cold inhomogeneities originate {\it in situ}, at
distances just prior to the formation of shocks, rather then resulting
from modulation of the outflow by a central engine.  Their formation
may be related to the transition from the Poynting flux to matter
dominated flow; then, the locally fast moving inhomogeneities can be a
product of intensive magnetic field reconnection events. Another
possibility is that the non-thermal flares are produced by
relativistic electrons directly accelerated in the reconnection sites.
However, this scenario can be questioned by the fact that the spectra
of injected electrons predicted theoretically and numerically are much
narrower than deduced from spectral analysis of the high energy
non-thermal flares \citep{zh01,llr03}.  In any case, given the
promising future of the MeV/GeV $\gamma$-ray astronomy with the
upcoming launch of the GLAST mission, this question will be addressed
directly by observations.  However, it is essential that simultaneous
observations in the soft X-ray band are conducted as well, although
the lack of a sensitive all-sky X-ray monitor will limit studies to
probably a limited number of pointed observations in the X-ray band by
satellites such as Chandra, XMM-Newton, and Astro-E2.

\acknowledgments This project was partially supported by Polish KBN
grants 5 P03D 00221 and PBZ-KBN-054/P03/2001, by Chandra grants
no. GO1-2113X and GO4-5125X 
from NASA via Smithsonian Astrophysical Observatory, and
by the Department of Energy contract to SLAC no. DE-AC3-76SF00515.
R.M. and M.S. thank SLAC for its hospitality during the completion of
part of this work. We thank the anonymous referee for her/his
valuable comments that helped to improve this paper.  

\appendix

\section{APPENDIX}

\subsection{Velocity of the shocked plasma and efficiency of the
energy dissipation}
\label{appA1}

Momentum and energy conservation for collision of two intrinsically
identical shells gives the velocity of the shocked plasma
\be
\beta = { \beta_1\Gamma_1 + \beta_2\Gamma_2 \over \beta_1 +\beta_2}
\, ,
\ee
and the efficiency of energy dissipation
\be
\eta_{diss} = 1 - {2 \Gamma \over \Gamma_1 + \Gamma_2} \, ,
\ee
where $\Gamma = 1/\sqrt{1-\beta^2}$.

For $\Gamma_2$ and $\Gamma_1 \gg 1$
\be
\beta \simeq 1 - {1 \over 2 \Gamma_1 \Gamma_2} \, ,
\ee
\be
\Gamma \simeq {1 \over \sqrt{1 -\beta^2}} \simeq \sqrt{\Gamma_1 \Gamma_2}
\, ,
\ee
and
\be
\eta_{diss} \simeq
{[(\Gamma_2/\Gamma_1)^{1/2}-1]^2 \over (\Gamma_2/\Gamma_1) + 1}\, .
\ee
%
%

For a given $\Gamma$ and $\alpha_{\Gamma} \equiv \Gamma_2/\Gamma_1$,
we have $\Gamma_1 \simeq \Gamma/\sqrt{\alpha_{\Gamma}}$ and $\Gamma_2
\simeq \Gamma \sqrt{\alpha_{\Gamma}}$.

\subsection{Shell velocities in the frame of the shocked plasma}
\label{appA2}

From Lorentz transformation of shell velocity from the central engine
frame to the shocked plasma frame
\be
\beta_i'  = {\beta_i - \beta \over 1 - \beta_i \beta} \, .
\ee
For $\Gamma_i$ and $\Gamma \gg 1$ it gives
\be
\beta_i' \simeq  {\Gamma_i^2 - \Gamma^2 \over \Gamma_i^2 + \Gamma^2}
\, ,
\ee
\be
\Gamma_i' \simeq {\Gamma_i^2 + \Gamma^2 \over 2 \Gamma_i \Gamma}
\, .
\ee
For $\Gamma \simeq \sqrt{\Gamma_1 \Gamma_2}$ the above formulae give
\be
\Gamma_1' = \Gamma_2' = {1 \over 2} {\alpha_{\Gamma} + 1 \over
\sqrt{\alpha_{\Gamma}}} \, ,
\ee
and
\be
\beta_2' = -\beta_1' = {\alpha_{\Gamma} -1 \over  \alpha_{\Gamma}+1}
\, .
\ee

%

\subsection{The shock width} 
\label{appA3}

From the shock theory, the ratio of the downstream density $n_{ds}$ to
the upstream density $n_{us}$ is
\be
{n_{ds} \over n_{us}} = {\hat \gamma \Gamma_{us}' +1 \over \hat \gamma-1}
\, ,
\ee
where $\Gamma_{us}'=\Gamma_{1,2}'$ is the bulk Lorentz factor of the
upstream flow as measured in the downstream (shocked plasma) frame,
and $5/3 < \hat \gamma < 4/3$ \citep[see][]{bk76}.  From
particle number continuity
\be
{\lambda_{0} \over \lambda} \simeq  {n_{ds} \over n_{us}}\, ,
\ee
where $\lambda_{0}$ is the proper width of the unshocked shell, and
$\lambda$ is the proper width of the shocked plasma layer.
In the downstream frame (corresponding to that of the 
shocked plasma/discontinuity surface), time
of the collision $t'_{coll}$ equals the time necessary for the shock
front to travel the distance $\lambda$
\be
t_{coll}' = {\lambda \over c \vert \beta_{s}' \vert } \, ,
\ee
where $\beta_{s}'$ is the velocity of the shock front.  In the shocked
plasma frame, the width of the unshocked plasma shell is
$\lambda_0/\Gamma'_{us}$ and thus during the collision, the outer
boundary of the unshocked plasma travels the distance
\be
\lambda_{0}/\Gamma_{us}' - \lambda = t_{coll}' c \vert \beta_{us}'
\vert \, ,
\ee
where $\vert \beta_{us}' \vert = \vert \beta_{1,2}' \vert$ is
the velocity of the upstream flow in the reference frame of the 
shocked plasma.
%
%

From above equations
\be
\vert \beta_{s}' \vert = \vert \beta_{us}' \vert
{(\hat \gamma -1) \Gamma_{us}' \over \Gamma_{us}' +1} \, ,
\ee
\be
\lambda = c t_{coll}' \vert \beta_s' \vert =
c t_{fl} {\cal D} g \, ,
\ee
and
\be
\lambda_{0} =  {\hat \gamma \Gamma_{us}' +1 \over \hat \gamma-1} \lambda
=c t_{fl} {\cal D}  g_0 \, , \label{lamb}
\ee
where 
\be
g={(\hat \gamma - 1) \vert \beta_{us}' \vert
\Gamma_{us}' \over \Gamma_{us}' +1 }\, ,
\ee
and 
\be
g_0={(\hat \gamma \Gamma_{us}' +1) \vert \beta_{us}' \vert
\Gamma_{us}' \over \Gamma_{us}' +1 }
\, .
\ee
Comparison of the exact value of the shell width given by
Eq. (\ref{lamb}) with its approximate value given by
Eq. (\ref{lambda}) gives
\be
{{\lambda_0^{(approx)} \over \lambda_0^{(exact)}}} = 
{\hat \gamma \Gamma_{us}' +1 \over \Gamma_{us}' +1} \,.
\label{ll}
\ee

\subsection{The pair content}
\label{appA4}

Amount of energy needed to be injected into relativistic electrons to
produce the observed non-thermal flares is
\be E_{e,inj}' = t_{coll}' L_{e,inj}' = 
t_{fl} {\cal D} \int_{\gamma_{min}}^{\gamma_{max}} Q m_e c^2 \gamma d\gamma
\, ,  \ee
where $Q$ is determined from the observed spectra (see
\S~\ref{numele}).  This energy cannot exceed the amount of energy
dissipated during the collision of two shells, which is
\be E_{diss}' = {(E_1 + E_2)\eta_{diss} \over \Gamma} \, ,  \ee
where for $n_p m_p \gg n_e m_e$
\be E_1 + E_2 \simeq N_p m_p c^2 (\Gamma_1 + \Gamma_2) \, .  \ee
Hence, for a given amount of electrons determined from observations of
non-thermal flares, the condition $ E_{diss}'>E_{e,inj}'$ gives the
minimum number of protons in shells, and, respectively, the maximum
pair content.  Parameterizing the fraction of dissipated energy tapped
by relativistic electrons by $\eta_e$, and noting that the number of
electrons per shell is $N_e = 0.5 N_{e,inj}= 0.5 t_{fl} {\cal D} \int
Qd\gamma$, we obtain
\be {N_e \over N_p} = \eta_e {m_p \over m_e} 
{(\sqrt{\Gamma_2/\Gamma_1} -1)^2 \over 2\sqrt{\Gamma_2/\Gamma_1} } 
{1 \over \bar \gamma_{inj}} \, ,  \ee
where $\bar \gamma_{inj} \equiv \int Q\gamma d\gamma/ \int Q d\gamma$
is the averaged energy of injected electrons. Since $n_e \equiv n_+ +
n_-$ and $n_p + n_+ = n_-$ (charge neutrality), the pair content
(number of pairs per proton) is $n_+/n_p = ((N_e/N_p) - 1)/2$.

\clearpage

\clearpage

\clearpage

\clearpage

\clearpage

\end{document}